\documentclass[12pt]{nature}

\usepackage{graphicx}
\usepackage{amssymb}
\usepackage{amsmath}

\title{Evidence of a one-dimensional thermodynamic phase diagram for simple glass-formers} 

\author{H.W. Hansen$^1$, A. Sanz$^1$, K. Adrjanowicz$^2$, B. Frick$^3$ \& K. Niss$^{1\ast}$}

\begin{document} 
\maketitle


\begin{affiliations}
 \item Glass and Time, IMFUFA, Department of Science and Environment, Roskilde University, Postbox 260, DK-4000 Roskilde, Denmark 
\item Institute of Physics, University of Silesia, ul. Uniwersytecka 4, 40-007 Katowice, Poland
\item Institut Laue-Langevin, 71 avenue des Martyrs,
CS 20156, 38042 Grenoble cedex 9, France
\end{affiliations}


\begin{abstract}
  The glass transition plays a central role in nature as well as
  in industry, ranging from biological systems such as proteins and DNA
  to polymers and metals \cite{Khodadadi15,Frick95,Luo17} . Yet the
  fundamental understanding of the glass transition which is a
  prerequisite for optimized application of glass formers is still
  lacking \cite{Albert16,Ediger12,Berthier11,Dyre06}.  Glass formers
  show motional processes over an extremely broad range of timescales,
  covering more than ten orders of magnitude, meaning that a full
  understanding of the glass transition needs to comprise this
  tremendous range in timescales \cite{Bengtzelius84,Buchenau92,Scopigno03}. 
  Here we report on first-time
  simultaneous neutron and dielectric spectroscopy investigations of
  three glass-forming liquids, probing in a single experiment the full
  range of dynamics. For two van der Waals liquids we locate in the
  pressure-temperature phase diagram lines of identical dynamics of
  the molecules on both second and picosecond timescales. This
  confirms predictions of the isomorph theory \cite{Dyre14} and effectively reduces
  the phase diagram from two to one dimension. The implication is that
  dynamics on widely different timescales are governed by the same
  underlying mechanisms.
\end{abstract}


Glasses are formed when the molecular motions of a liquid become so
slow that it effectively becomes a solid. When the glass transition of
a liquid is approached, the dynamics of the molecules spreads out like
a folding fan covering more than ten orders of magnitude. There are at
least three overall contributions to the dynamics of glass-forming
liquids: 1) vibrations, 2) fast relaxations on picosecond timescales, and 3) the structural alpha relaxation, which has a strongly
temperature-dependent timescale. The glass transition occurs when the
alpha relaxation is on a timescale of hundreds of seconds and
therefore completely separated from the two fast contributions to
dynamics. Nonetheless, both theoretical \cite{Bengtzelius84,Gotze09} and experimental results \cite{Buchenau92,Sokolov93,Scopigno03,Novikov04,Larini08} have suggested that fast and slow dynamics are intimately connected.  A complete understanding of the glass transition therefore necessitates a full understanding of all these
dynamic processes.

During the past 15 years, pressure has increasingly been introduced to
study dynamics of glass-forming liquids in order to disentangle
thermal and density contributions to the dynamics. The most striking
finding from high-pressure studies is that the alpha relaxation, both
its timescale and spectral shape is, for a large number of different
liquids, uniquely given by the parameter $\Gamma=\rho^\gamma/T$,
where $\gamma$ is a material specific constant
\cite{Tolle98,Tarjus04,Roland05,Casalini14_dTdP}. This scaling
behaviour can be explained by the isomorph theory
\cite{Gnan09,Dyre14}, which, moreover, predicts the value of $\gamma$
(\citeonline{Gundermann11}). The fundamental claim of isomorph
theory is the existence of isomorphs. Isomorphs are curves in the
phase diagram along which \emph{all} dynamic processes and structure
are invariant.  Put
in other words, the phase diagram is predicted to be one dimensional
with respect to structure and dynamics on all timescales, with the
governing single variable being $\Gamma$. Isomorph theory has been very successful
in describing Lennard-Jones type computer simulated liquids,
e.g. refs. (\citeonline{Bacher14,Pedersen16}). Experimental studies
of isomorph theory predictions require high-precision high-pressure measurements and 
are still limited
\cite{Gundermann11,Xiao15,Adrjanowicz16,Romanini17}. Consequently, it remains open whether isomorph theory holds for real molecular
liquids.

The only systems that obey isomorph theory exactly are those with
repulsive power law interaction potentials which do not, of course,
describe real systems. Hence, isomorph theory is approximate in
its nature and expected to work for systems without directional
bonding and competing interactions \cite{Dyre14}.  With this in mind,
we have studied the dynamics on three well-studied glass formers
representing non-associated liquids and liquids with directional
bonding, two van der Waals bonded liquids (vdW-liquids): PPE (5-polyphenyl ether) and
cumene (isopropyl benzene), and a hydrogen bonding (H-bonding) liquid: DPG
(dipropylene glycol).

In this work, we use a new high-pressure cell for simultaneous
measurements of the fast dynamics by neutron spectroscopy and the
alpha relaxation by dielectric spectroscopy to demonstrate that for the studied vdW-liquids, the three mentioned distinct dynamic components are invariant along the same lines in the phase
diagram. This is the first direct experimental evidence for the
existence of isomorphs, and it proves that the phase diagram of simple vdW-liquids is one dimensional with respect to dynamics. Unlike the scaling behaviour of the alpha relaxation dynamics, which is often found to hold surprisingly well in H-bonding systems \cite{Adrjanowicz16,Puosi16,Romanini17}, we only find invariance of the fast relaxational and vibrational dynamics on picosecond timescales in the vdW-liquids. For the single investigated H-bonding system, DPG, we make a different observation, as expected based on isomorph theory.

Dynamics from picosecond to kilosecond cannot be measured with one
single technique; several complementary techniques are required. A
glass-forming liquid is in metastable equilibrium and the dynamics is
very sensitive to even small differences in pressure and
temperature. The high viscosity of the liquid close to the glass
transition temperature, $T_g$, makes the transmission of isotropic
pressure non-trivial, as pressure gradients are easily generated. In
order to ensure that the different dynamics are measured under
identical conditions, we have therefore developed a cell for doing
simultaneous dielectric spectroscopy (DS) and neutron spectroscopy
(NS) under high pressure (Fig.~1). The experiments
were carried out on spectrometers at the Institut Laue-Langevin (ILL) on the
time-of-flight (TOF) instruments IN5
and IN6. The different NS instruments access different timescales with IN5 giving information on the $\sim$10~ps scale, while a backscattering (BS) instrument like
IN16 accesses $\sim$1~ns dynamics. DS provides fast
(minutes) and high accuracy measurements of the dynamics from
microsecond to 100~s.

The dynamics measured with the different techniques are illustrated in
Fig.~1a~and~b for PPE. The center panels of (a) and (b) are sketches of the incoherent
intermediate scattering function, $I(Q,t)$, while the top and bottom panel show raw
data. At $T_g$ (Fig.~1a), no broadening is observed on nanosecond timescales (IN16) corresponding to a plateau in $I(Q,t)$, on picosecond timescales from IN5 we observe contributions from fast
relaxational processes and vibrations, whereas the alpha relaxation is
seen in DS at much longer timescales, a difference of more than 10
orders of magnitude. As the temperature is increased, the processes
merge (Fig.~1b), and relaxation dominates the signal in all three
spectrometers.

Picosecond dynamics measured on IN5 and IN6 are presented in Fig.~2. We observe
the same trend for all spectra for all values of wave vector $Q$ (Extended Data Fig.~2), and have summed over $Q$ to improve statistics. All spectra are shown on the same $S(\tilde{\omega})$-axis. Motivated by isomorph theory, the energy scale is shown in reduced
units, effectively $\tilde{\omega} = \omega\rho^{-1/3}T^{-1/2}$
\cite{Dyre14}. The effect of scaling is small, though
visible, in the studied range of $\rho$ and $T$. The data is shown on an absolute energy scale in the Extended Data Fig.~1.

For all three samples in row (a) Fig.~2, which shows dynamics in the
liquid, we observe the extreme scenarios sketched in Fig.~1. At low
pressure, relaxational contributions are dominating (Fig.~1b). At the
glass transition, we find only fast relaxational and vibrational
contributions (Fig.~1a).  The fast relaxational contributions decrease
in the glassy state (Fig.~2b), leaving the excess
vibrational density of states, which shows up as the so-called Boson
peak \cite{Chumakov11}, as the dominant contribution (black full lines in Fig.~2b). For
all three samples, we observe different dependencies on temperature
and pressure for the three contributions to the dynamics, such that
their relative contributions vary along both isobars and isotherms.

Sokolov et al. \cite{Hong09} studied seven different liquids with
Raman spectroscopy, both H-bonding and vdW-liquids. In contrast to what we see, they observe a correlation between
pressure-induced variations in the fast relaxation and the Boson
peak. Our observations are in agreement with those shown by neutron
spectroscopy for ortho-terphenyl in (\citeonline{Patkowski03}), namely that
the $T$ and $P$ dependencies are different for the three
contributions to the dynamics.
 
All glass formers have isochrones which are lines in the $(T,P)$-phase
diagram with constant alpha relaxation time, $\tau_\alpha$. If
isomorphs exist in a liquid, these coincide with the isochrones, since all dynamic processes on all timescales and of all dynamic
variables are invariant along an isomorph. Thus, experimentally we can
identify candidates for isomorphs by the isochrones. We use DS effectively
as a 'clock', which identifies the alpha relaxation time from the
dipole-dipole correlation function, while simultaneously measuring the
picosecond dynamics with incoherent neutron scattering probing the particle self-correlation.

Row (c) in Fig.~2 shows the picosecond dynamics measured along the
glass transition isochrone $T_g(P)$ defined as when
$\tau_\alpha=100$~s found from DS. We observe superposition of the
spectra at the picosecond timescale, thus invariance of the dynamics,
for the two vdW-liquids (PPE and cumene). This is in agreement with
the prediction of the isomorph theory and it is particularly striking
because at $T_g(P)$, fast relaxational and vibrational motion are
completely separated in timescale from the alpha relaxation as
illustrated in Fig.~1a. In contrast, for the H-bonding liquid (DPG)
we find a clear shift towards higher energy and an intensity decrease
of the Boson peak along the $T_g(P)$ isochrone. The lack of
superposition in the H-bonding system demonstrates that the
superposition seen in the vdW-liquids is non-trivial. Thus, the
superposition observed in the vdW-liquids is a genuine signature of the isomorphs in these liquids.

To compare more state points, including shorter alpha-relaxation time
isochrones ($\tau_\alpha<1~\mu$s), isotherms and isobars, we plot the
corresponding inelastic intensities at a fixed reduced energy
($\tilde{\omega}=0.06$) as a function of temperature and pressure
(Fig.3a and b). Along the investigated isochrones the inelastic intensity at
$\tilde{\omega}=0.06$ or $t\sim 1$~ps is found to be invariant for the
vdW-liquids. Again the H-bonding liquid behaves differently and its
picosecond dynamics varies along the isochrones.  
This confirms the isomorph prediction of spectral superposition of
fast relaxation, vibrations/Boson peak and alpha relaxation for the
vdW-liquids: both when there is timescale separation and
when the processes are merged.

The power of isomorph theory is that the phase diagram becomes one
dimensional; the dynamics only depend on which isomorph a state point
is on and not explicitly on temperature and pressure (or
density). Since isomorphs are isochrones the other dynamic
components should become a unique function of the alpha relaxation
time. In Fig.~3(c) we plot the intensity of the picosecond dynamics
along isotherms and isobars as a function of the dielectric alpha
relaxation time at fixed reduced energies
($\tilde{\omega}=0.02,~0.04,~0.06,~0.08,~0.1$). The data from each
energy collapses in this plot, illustrating that, as predicted, all the dynamics is
governed by one parameter.

Previously suggested connections between fast and slow dynamics,
e.g. via the temperature dependence of properties
\cite{Dyre06,Larini08}, often suggest a causality, where the fast
dynamics controls the slow dynamics.  In contrast, the connection between slow and fast dynamics shown in this work does not tell us if one controls the other or if they are simply controlled by the same underlying mechanism.

Isomorph theory is approximate in its nature, and the isomorphs of
real physical systems are approximate. The fact that the isomorph
prediction works for dynamics which is separated in timescale by more
than ten orders of magnitude, tells us that whatever governs this
dynamics is controlled by properties of the liquid that obey the
isomorph scale invariance. There is a consensus that the alpha
relaxation is cooperative, fast relaxations are normally
understood as cage rattling, whereas it has been heavily debated
whether or not the Boson peak is localized \cite{Chumakov11,Hong09,Wyart10,Ferrante13}. It is
clear from the spectra in Fig.~2 that fast relaxation and Boson peak
are different in nature because their relative intensities vary along isotherms and isobars, and that these two types of fast dynamics are
distinctively different from the structural alpha relaxation. Yet all
of these dynamic features are controlled by the single parameter $\Gamma=\rho^\gamma/T$. Our finding implies that a universal theory for the glass-transition needs to be consistent with the one-dimensional phase diagram coming out of isomorph theory.




\begin{addendum}
 \item This work was supported by the Danish Council
for Independent Research (Sapere Aude: Starting Grant). 
We gratefully acknowledge J. Dyre for fruitful discussions and technical support from the workshop at IMFUFA and the SANE group at the ILL.
 \item[Competing Interests] The authors declare that they have no
competing financial interests.
 \item[Correspondence] Correspondence and requests for materials
should be addressed to K. Niss.~(email: kniss@ruc.dk).
\end{addendum}


\newpage

\topmargin 0.0cm
\oddsidemargin 0.2cm
\textwidth 16cm 
\textheight 21cm
\footskip 1.0cm

\baselineskip24pt

\begin{figure}[hbtp!]
\centering
\includegraphics[width=12 cm]{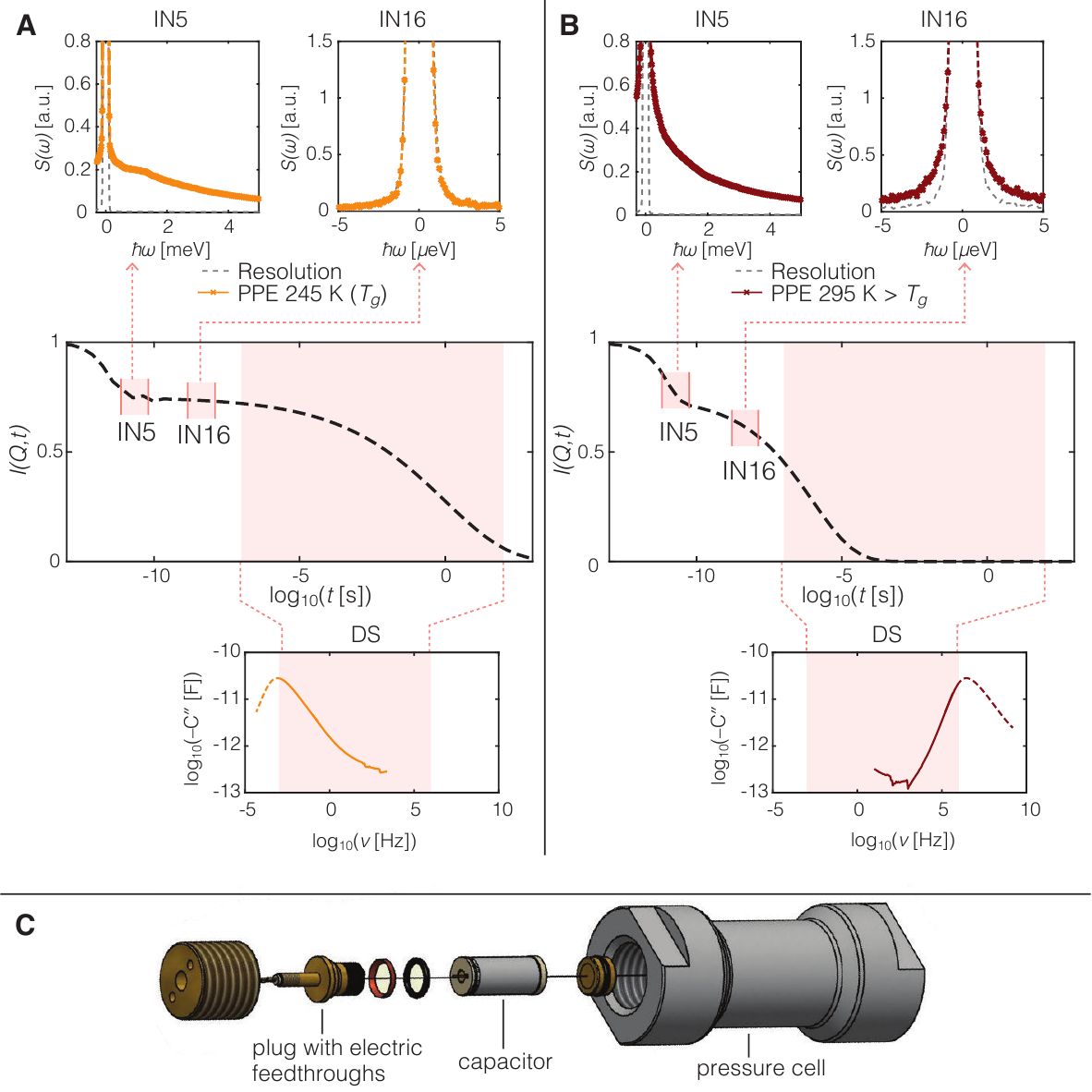}
\caption{\small Dynamics probed with dielectric
  (DS) and neutron spectroscopy. (a) and (b) sketch the incoherent
  intermediate scattering function $I(Q,t)$ in the center panel. Top 
  panels show raw spectra measured on the TOF and BS spectrometers IN5 and IN16, respectively, and bottom panels show spectra from DS, all from
  PPE. Dashed lines in DS are from time-temperature
  superpositon (TTS). (a) Dynamics at the glass transition $T_g$ where
  there is separation of timescales. At picosecond timescales we
  observe fast relaxation and vibrations. At nanosecond timescales
  there is no relaxation and the alpha relaxation is only visible in
  DS. (b) Dynamics in the liquid well above $T_g$. Relaxation
  dominates the signal in all three spectrometers. (c) Drawing of
  components of simultaneous high-pressure dielectric and neutron
  spectroscopy cell.}
\end{figure}

\begin{figure}[hbtp!]
\centering
\includegraphics[width=12cm]{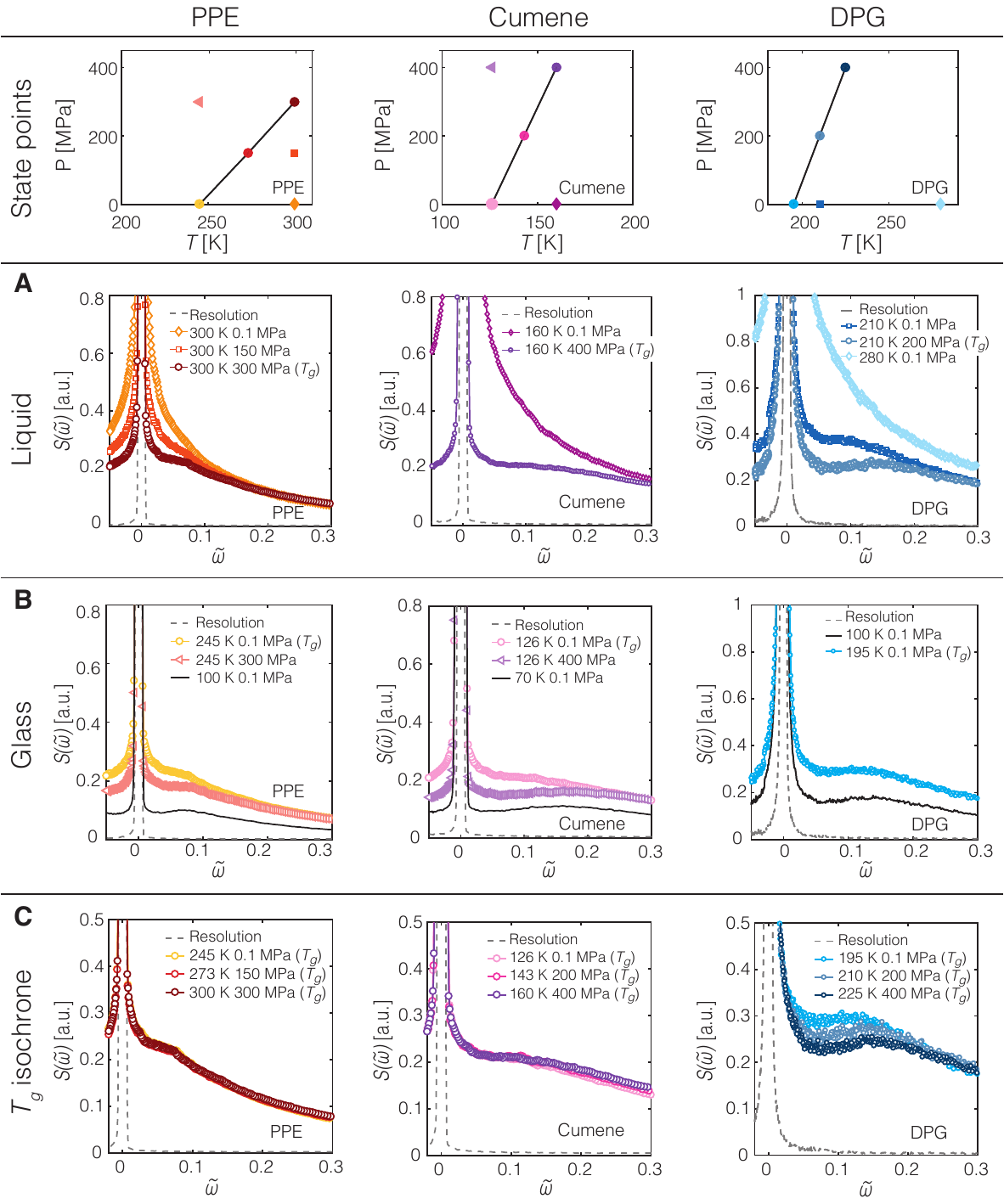}
\caption{\small Picosecond dynamics at various state points for PPE, cumene
  and DPG. Top panel: $(T,P)$-phase diagram showing the state points
  of the spectra in row (a-c). The black lines correspond to the glass
  transition isochrones with $\tau_\alpha=100$~s. All spectra are
  plotted in reduced units, $\tilde{\omega}=\omega\rho^{-1/3}T^{-1/2}$
  and summed over $Q$. (a) Spectra in the liquid: at
  low pressure we primarily observe relaxation. As pressure is increased,
  reaching the glass transition, only fast relaxation and vibrations
  are left. (b) Spectra in the glass: leaving the glass transition,
  going deeper into the glass, fast relaxations disappear and only
  vibrations are left. (c) Along the glass transition isochrones there
  is superposition of dynamics for the two vdW-liquids, PPE
  and cumene, while in the H-bonding liquid, DPG, there is a
  clear shift in data with pressure. PPE and cumene were measured on IN5 and DPG on IN6. No scaling has been done on the $y$-axis, i.e. 
  all spectra are shown on the same $S(\tilde{\omega})$ scale.}
\end{figure}	

\begin{figure}[hbtp!]
\centering
\includegraphics[width=0.5\textwidth]{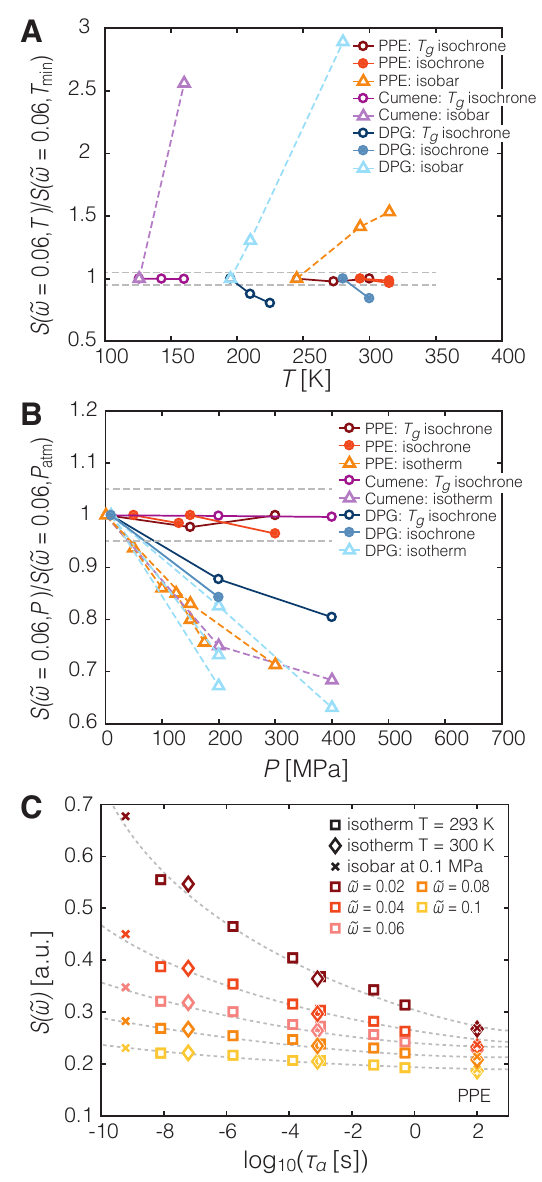}
\caption{\small Comparison of state points via the inelastic intensity at
  fixed reduced energies. (a) and (b) Inelastic intensity at fixed
  reduced energy ($\tilde{\omega}=0.06$) along isochrones, isotherms
  and isobars from the spectra in Fig.~2 plotted (a) as a function of
  temperature normalized to value at lowest temperature, (b) as a
  function of pressure normalized to the value at ambient
  pressure. (c) Inelastic intensity for fixed reduced energies for PPE
  as a function of $\tau_\alpha(T,P)$ for two isotherms, 293~K ({\tiny
    $\square$}), 300~K ($\diamond$) and an isobar ({\footnotesize
    $\times$}) for the reduced energies
  $\tilde{\omega}=0.02,~0.04,~0.06,~0.08,~0.1$ covering more than ten
  orders of magnitude in relaxation time. All lines are guides for the
  eye.}
\end{figure}


\newpage

\section*{Methods section for \\ "Evidence of a one-dimensional thermodynamic phase diagram for simple glass-formers"} 

\subsection*{H.W. Hansen$^1$, A. Sanz$^1$, K. Adrjanowicz$^2$, B. Frick$^3$ \& K. Niss$^{1\ast}$}


\subsubsection*{Materials}

Isopropyl benzene (cumene) and dipropylene glycol (DPG) were purchased from Sigma Aldrich, and 5-polyphenyl ether (PPE) was
purchased from Santolubes. All three samples were used as acquired.

The glass transition for the three samples is found from dielectric spectroscopy and defined as when the maximum of the loss peak corresponds to $\tau_\alpha=100$~s, where $\tau_\alpha=1/2\pi\nu''_\mathrm{max}$. At atmospheric pressure, the glass transition temperature $T_g$ is 245~K for PPE, 126~K for cumene, and 195~K for DPG.

The traditional way of quantifying how much the alpha relaxation time, or the viscosity, as a function of temperature deviates from Arrhenius behavior is given by the fragility \cite{Angell91}, defined as:
\begin{equation}
m = \frac{\mathrm{d}\log_{10}\tau_\alpha}{\mathrm{d}(T_g/T)}\biggr\rvert_{T_g}.
\end{equation}
For the two van der Waals liquids, the fragility at ambient pressure is $m\approx80$ for PPE and $m\approx70$ for cumene. The hydrogen-bonding liquid DPG has fragility $m\approx60$. 

\subsubsection*{Methods}

All experiments were carried out at the Institut
Laue-Langevin (ILL) on the time-of-flight instruments IN5 and IN6. In neutron spectroscopy, the different instrumental energy resolutions give access to different dynamical timescales; the coarser the energy resolution the faster the time window accessible: $\Delta E_\mathrm{res}\approx$~0.1~meV on IN5 and IN6 corresponds to $\sim$10~ps. DS provides fast (minutes) and high accuracy measurements of the dynamics from microsecond to 100~s. Details on the high-pressure cell for doing simultaneous dielectric and neutron spectroscopy can be found in the forthcoming publication.

In Fig.~2, we presented data on PPE and cumene from IN5 and on DPG from IN6 (hence the difference in statistics). All data was measured with a wavelength of 5~{\AA} and an energy resolution $\sim$0.1~meV, corresponding to a timescale of approximately 10 picoseconds. All spectra have been corrected in the conventional way by normalizing to monitor and vanadium, subtracting background, and correcting for self-shielding, self-absorption and detector efficiency using LAMP, a data treatment program developed at the ILL. The data has then been grouped for constant wavevector $Q$ in steps of 0.1~{\AA}$^{-1}$ in the range 1.2-1.9~{\AA}$^{-1}$ for the IN5 data (PPE and cumene) and in the range 1.2-1.7~{\AA}$^{-1}$ for the IN6 data (DPG) and is presented in Fig.~2 as a sum over $Q$. Data are shown as measured in Extended Data Fig.~ED1 on an absolute energy scale. No scaling has been done on the $y-$axis in $S(\omega)$ of any of the spectra, i.e. all spectra are plotted on the same scale. Comparing Fig.~S1 to Fig.~2, the effect of plotting data in reduced units is visible at higher energy transfer. The same picture is observed for all spectra for each value of $Q$, and we have therefore summed over $Q$ in the data shown in Fig.~2 to improve statistics. An example of spectra along the glass transition at different $Q$ for PPE and DPG is shown in Extended Data Fig.~ED2.

\subsubsection*{Reduced units}

According to isomorph theory, the relevant scale to look at is in reduced units $^{11}$. The reduced energy units used in Fig.~2 and Extended Data Fig.~ED2 are given by
\begin{equation}
\tilde{\omega} = \omega t_0 = \omega\rho^{-1/3}\sqrt{m/(k_BT)}
\end{equation} 
where $\omega$ is the energy transfer, setting $\hbar=1$. $k_B$ is Boltzmann's constant and $T$ is temperature. Here, $\rho$ is the number density and $m$ is the average particle mass, the latter assumed constant. We set $m=k_B=1$. Effectively, this becomes 
\begin{equation}
\tilde{\omega} = \omega\rho^{-1/3}T^{-1/2},
\end{equation}
where $\rho$ is now the volumetric mass density.

Just like reduced energy units, wave vector or momentum transfer $Q$ should also be presented in reduced units: 
\begin{equation}
\tilde{Q} = Q \rho^{-1/3}
\end{equation} 
But as the density changes are in the percent range in this study, scaling of $Q$ will be around 1\% and will be within the uncertainty of the data and is therefore neglected. 

\subsubsection*{Calculating density}

Equations of state (EOS) have been used to calculate the temperature and pressure dependence of the density. For DPG, the EOS is taken from \cite{Grzybowski11}. For PPE, a fit to the Tait equation from PVT data from \cite{Gundermann13_thesis} has been used to obtain the density:
\begin{equation}
\rho(T,P) = \left(V_0\exp(\alpha_0T)\left\lbrace1-C\ln\left[1+\frac{P}{b_0\exp(-b_1T)}\right]\right\rbrace\right)^{-1},
\end{equation}
where $\rho$ is in g/cm$^3$ and equal to $1/V_{sp}$, the specific volume, $P$ is pressure in MPa and $T$ is temperature in $^\circ$C. The fitting parameters are $V_0=0.82$, $\alpha_0=6.5\cdot10^{-4}$, $C=9.4\cdot10^{-2}$, $b_0=286$ and $b_1=4.4\cdot10^{-3}$. For cumene, density has been reported in \cite{Barlow66} in the temperature range $150-320$~K at atmospheric pressure, a linear dependence was found in this range encapsulated in the equation,
\begin{equation}
\frac{\rho}{\rho_r}=1-a(T-T_r),
\end{equation}
where $a$ is a constant given for cumene as $a=0.000954$~K$^{-1}$, and $\rho_r$ is the density at a reference temperature of $T_r=273.2$~K, $\rho_r=0.879$. This linearity is assumed to hold to  the glass transition temperature at ambient pressure, 126~K. The pressure dependence of density was measured in \cite{Bridgman49} up to 4~GPa at room temperature. To obtain an EOS, the compressibility and expansivity in the whole temperature range is found using the approach from \cite{Alba88} for toluene. The pressure and temperature dependence of the $\alpha(P)$ is calculated at temperatures higher than 240~K using the formula in \cite{Alba88} for toluene rescaled to cumene by using its critical temperature and density, $T_c=631.1$~K and $P_c=321$~kPa \cite{Wilson96}.

For all three samples, the density changes are in the percent range in the temperature-pressure range of this study. The PVT data and EOS are only used for scaling on the energy axis of the data, where the reduced energy units contain the cubic root of the density. Hence, the scaling is in practice mainly with temperature, and the use of EOS therefore does not alter with the overall conclusion.

\subsubsection*{Higher temperature isochrones}

High temperature isochrones with alpha relaxation time $\tau_\alpha\ll100$~s shown in Fig.~3 were done for PPE and DPG, but not for cumene; because cumene crystallizes easily in the region above $T_g$ and below the melting point. 

In Extended Data Fig.~ED3~and~ED4, examples of higher temperature isochrones, where relaxation is dominating in the picosecond dynamics, are given for PPE. In Extended Data Fig.~ED3, the imaginary part of the capacitance used to find isochrones are shown, both at the glass transition and for faster relaxation times. Extended Data Fig.~ED4 shows the corresponding NS spectra to the two high-temperature isochrones shown in Extended Data Fig.~ED3. We observe total invariance of the spectra along isochrones. Again, no scaling has been done in the $y$-direction, and in this interval, the effect of plotting in reduced units on the $x$-axis is negligible.


\newpage

\setcounter{figure}{0}
\renewcommand{\thefigure}{ED\arabic{figure}} 
\newcommand{\plotpath}{plots}

\begin{figure}
\begin{center}
\includegraphics[width=0.3\textwidth]{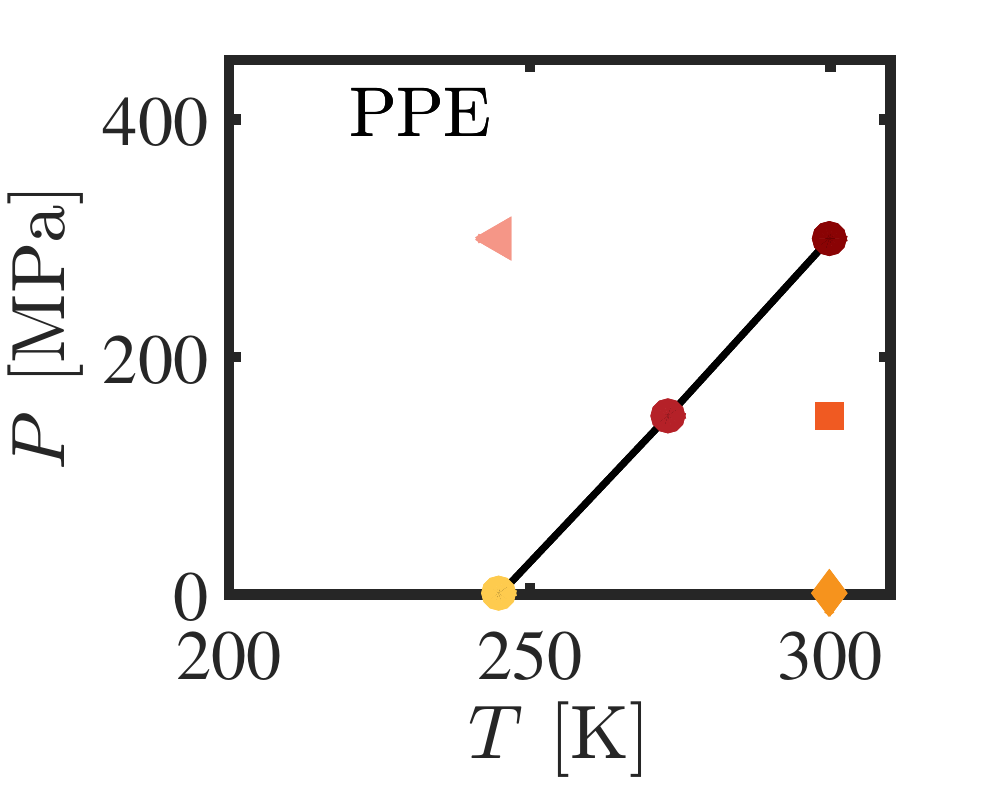}
\hspace{0.1em}
\includegraphics[width=0.3\textwidth]{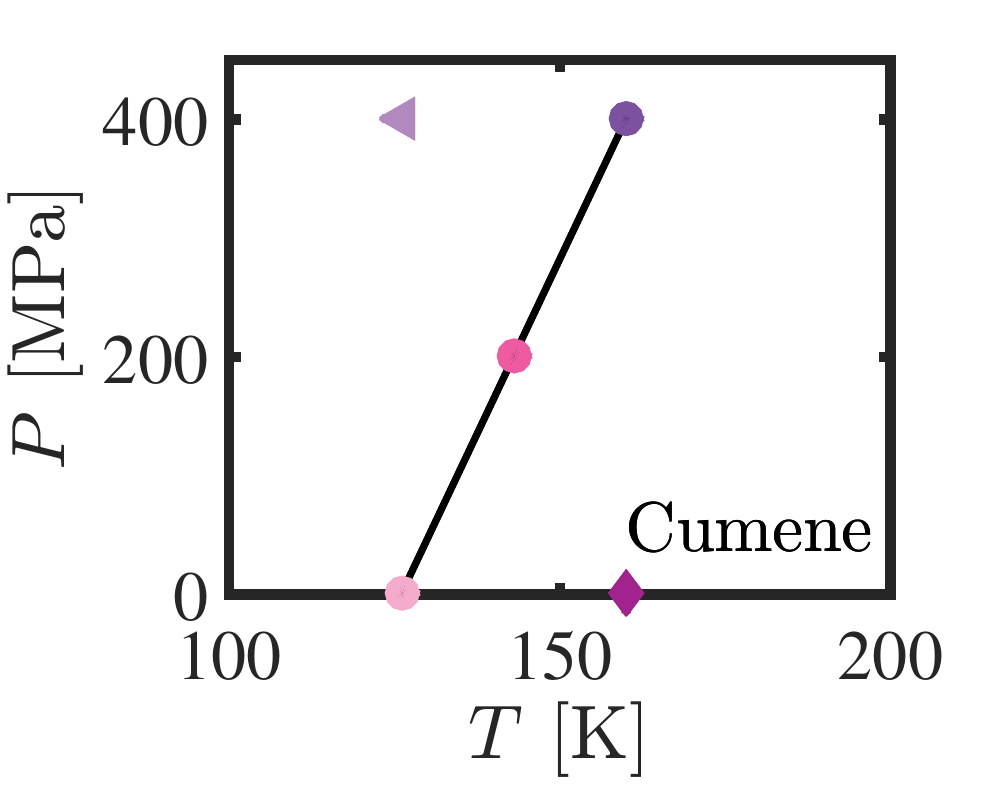}
\hspace{0.1em}
\includegraphics[width=0.3\textwidth]{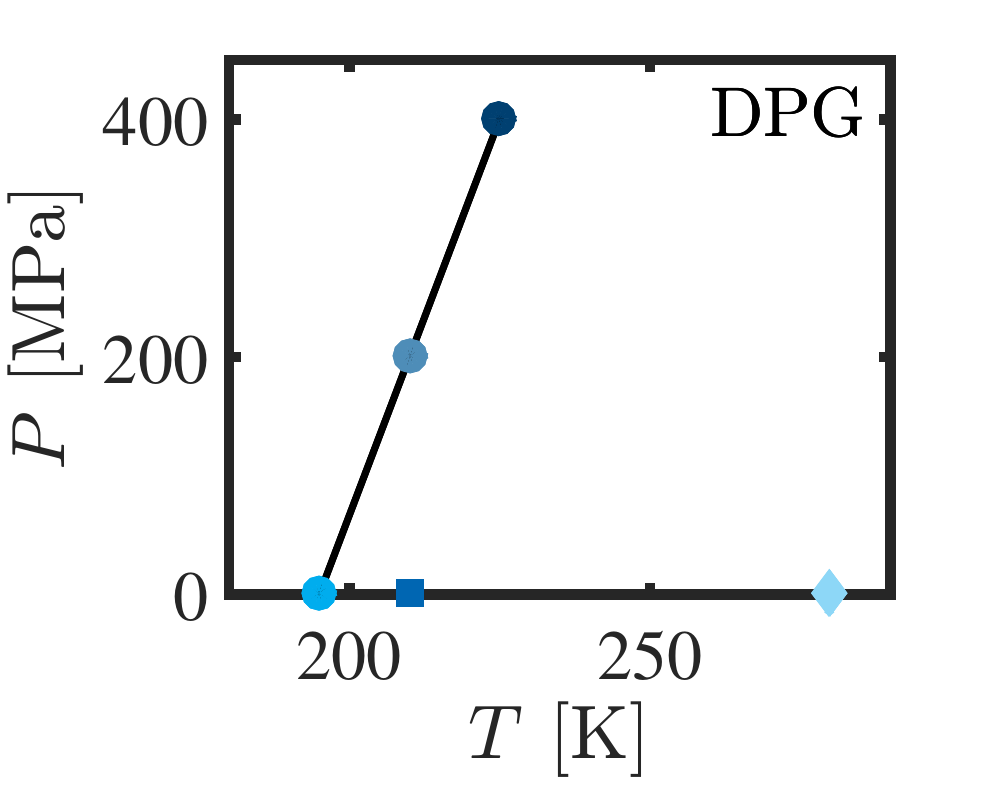} 
\end{center}
\textbf{a}
\vspace{-3em}
\begin{center}
\includegraphics[width=0.3\textwidth]{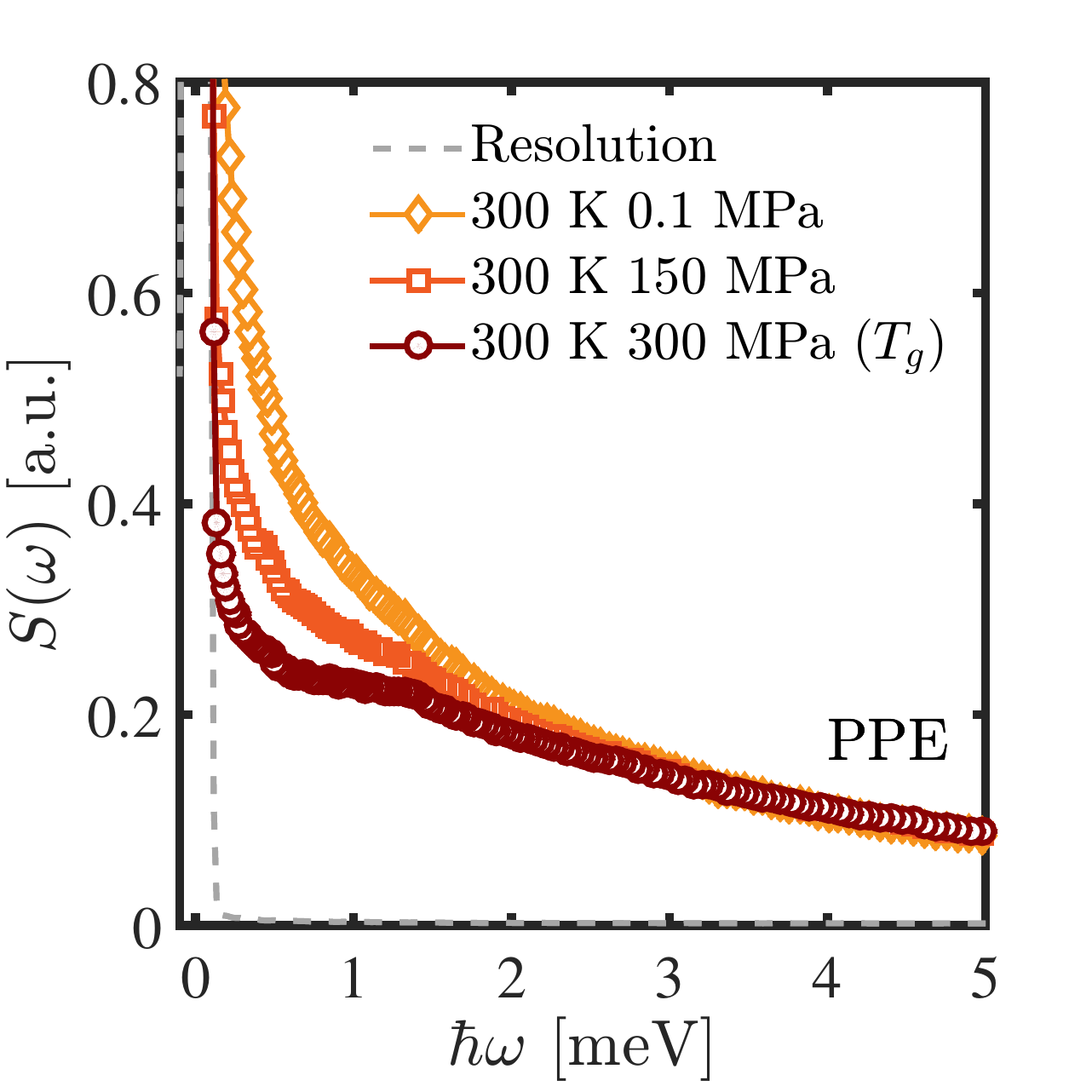} \hspace{0.1em}
\hspace{0.1em}
\includegraphics[width=0.3\textwidth]{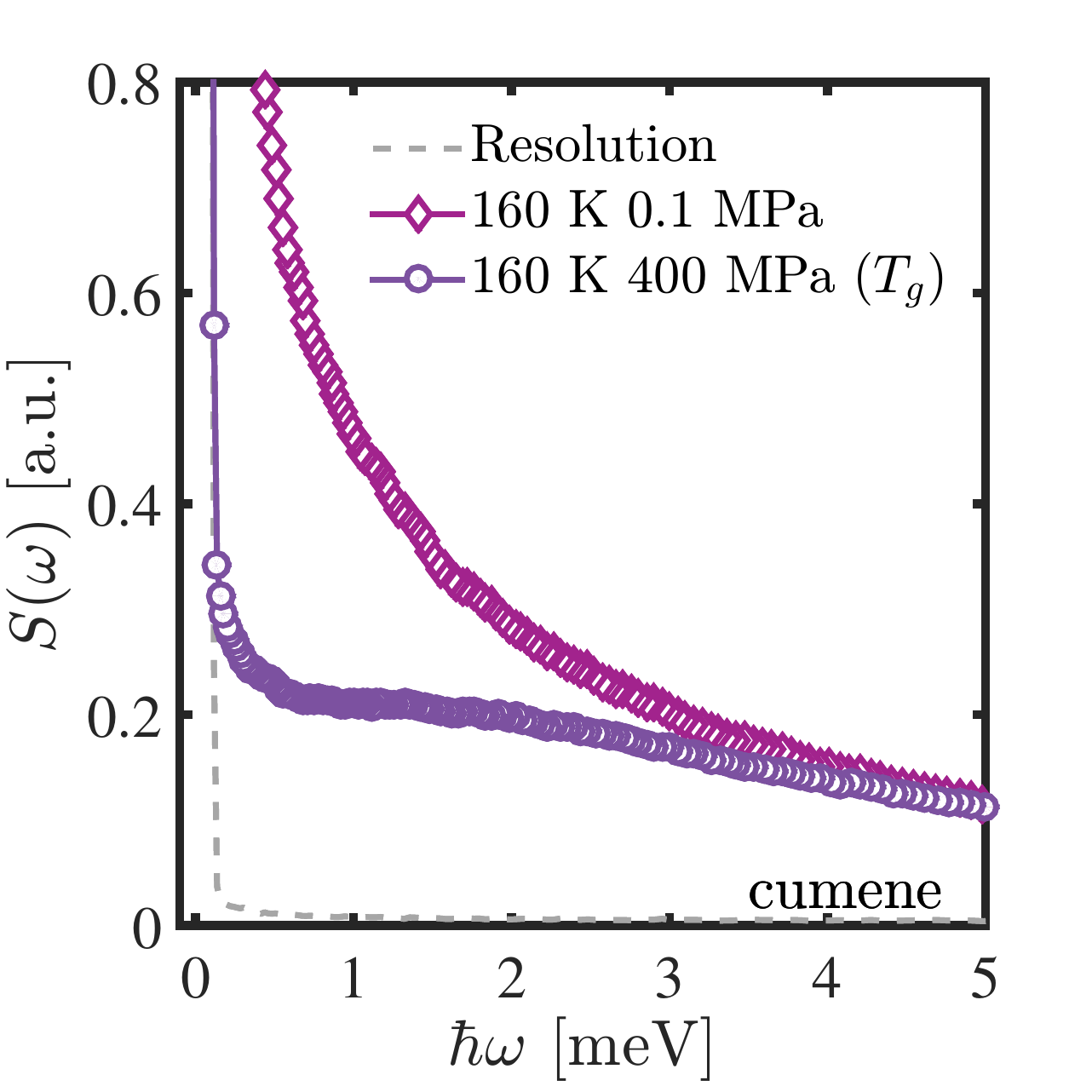}
\hspace{0.1em}
\includegraphics[width=0.3\textwidth]{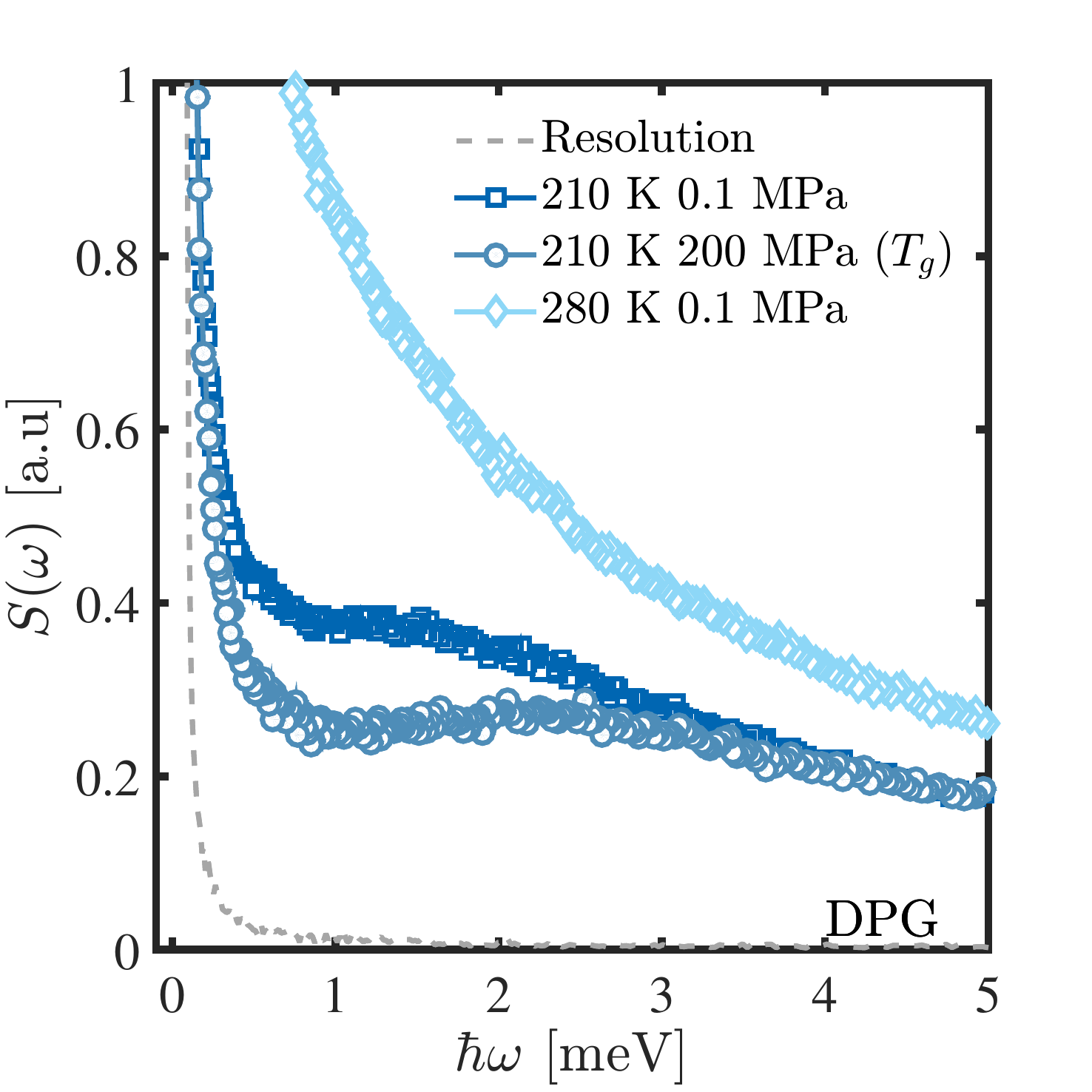} 
\end{center}
\textbf{b}
\vspace{-3em}
\begin{center}
\includegraphics[width=0.3\textwidth]{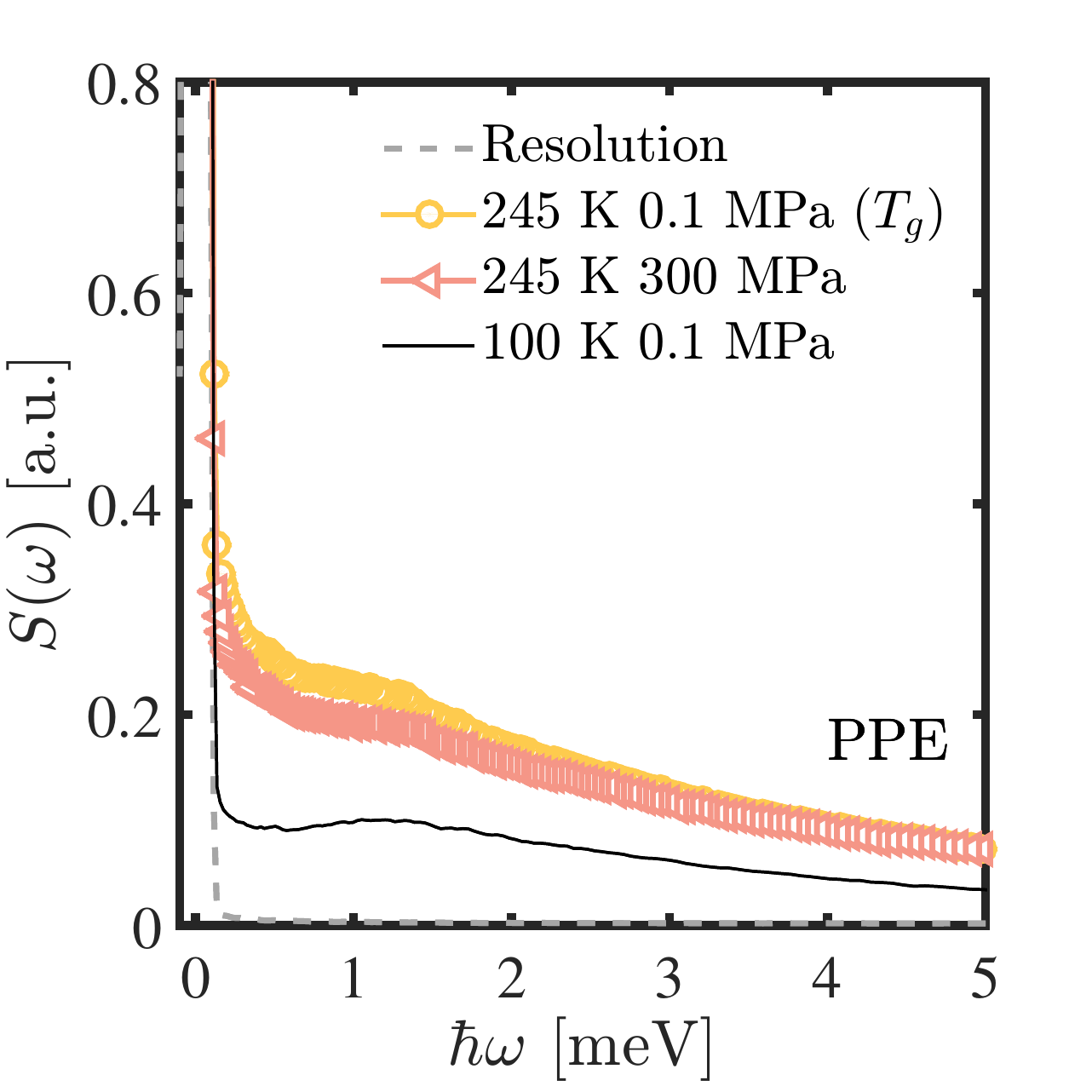} 
\hspace{0.1em}
\includegraphics[width=0.3\textwidth]{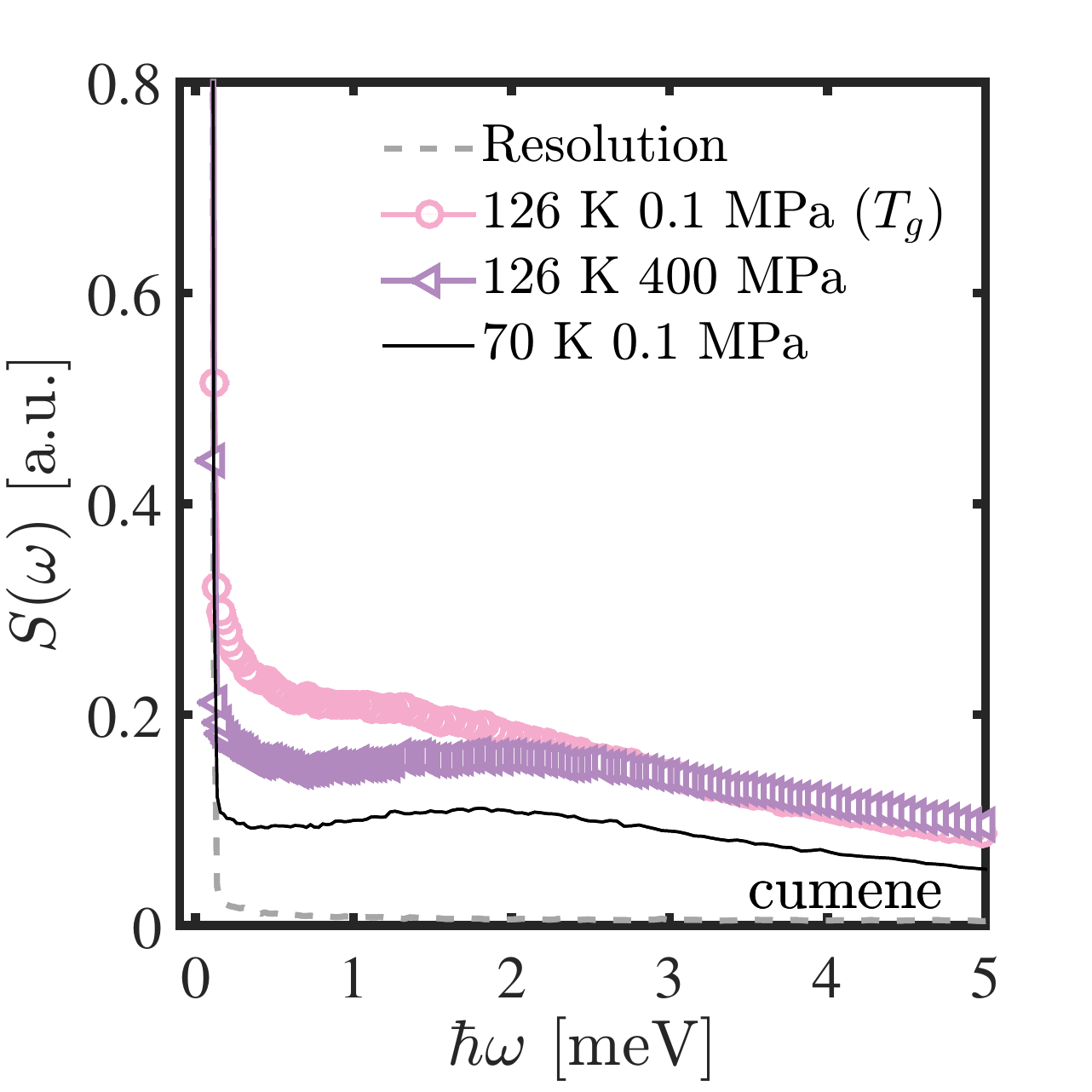}
\hspace{0.1em}
\includegraphics[width=0.3\textwidth]{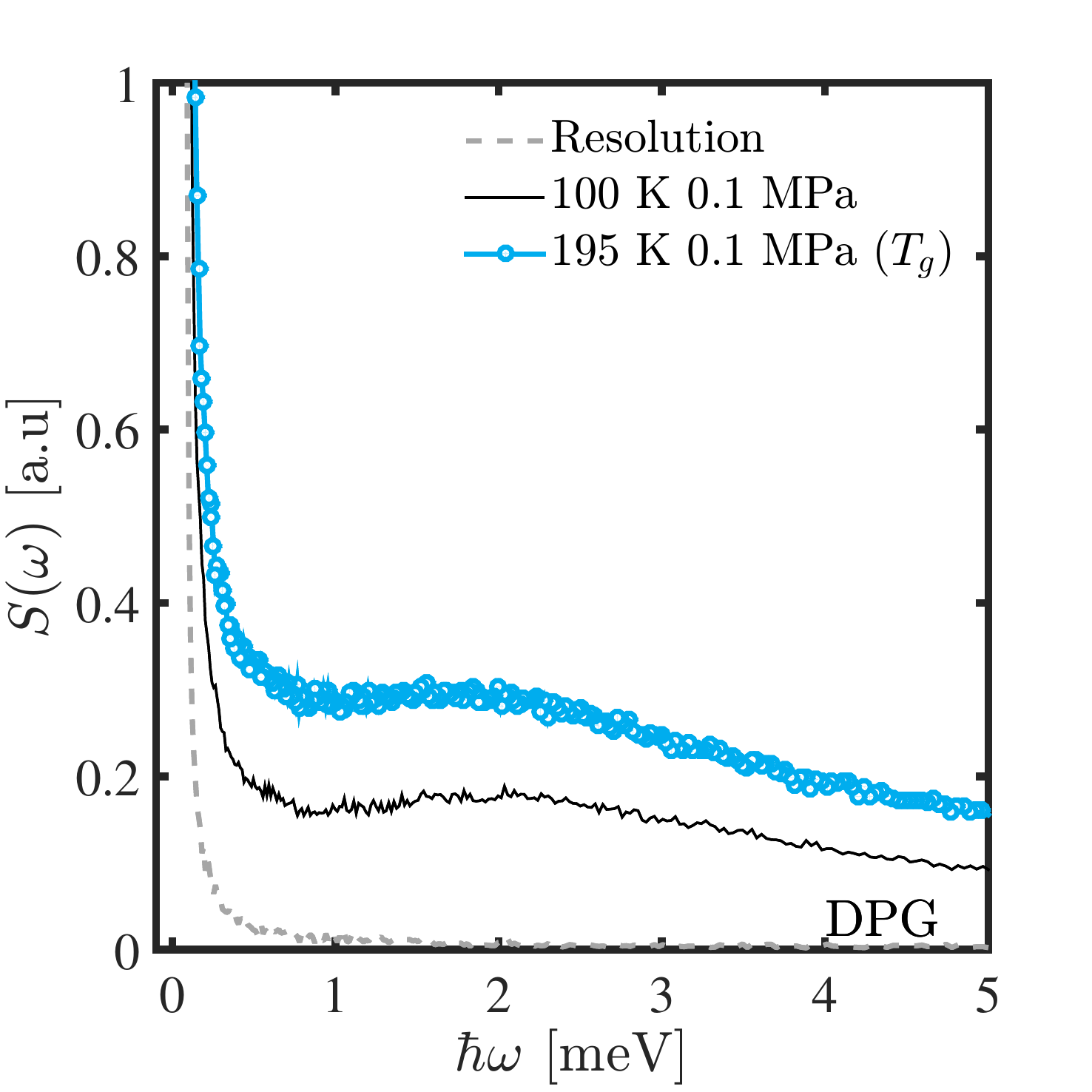} 
\end{center}
\textbf{c}
\vspace{-3em}
\begin{center}
\includegraphics[width=0.3\textwidth]{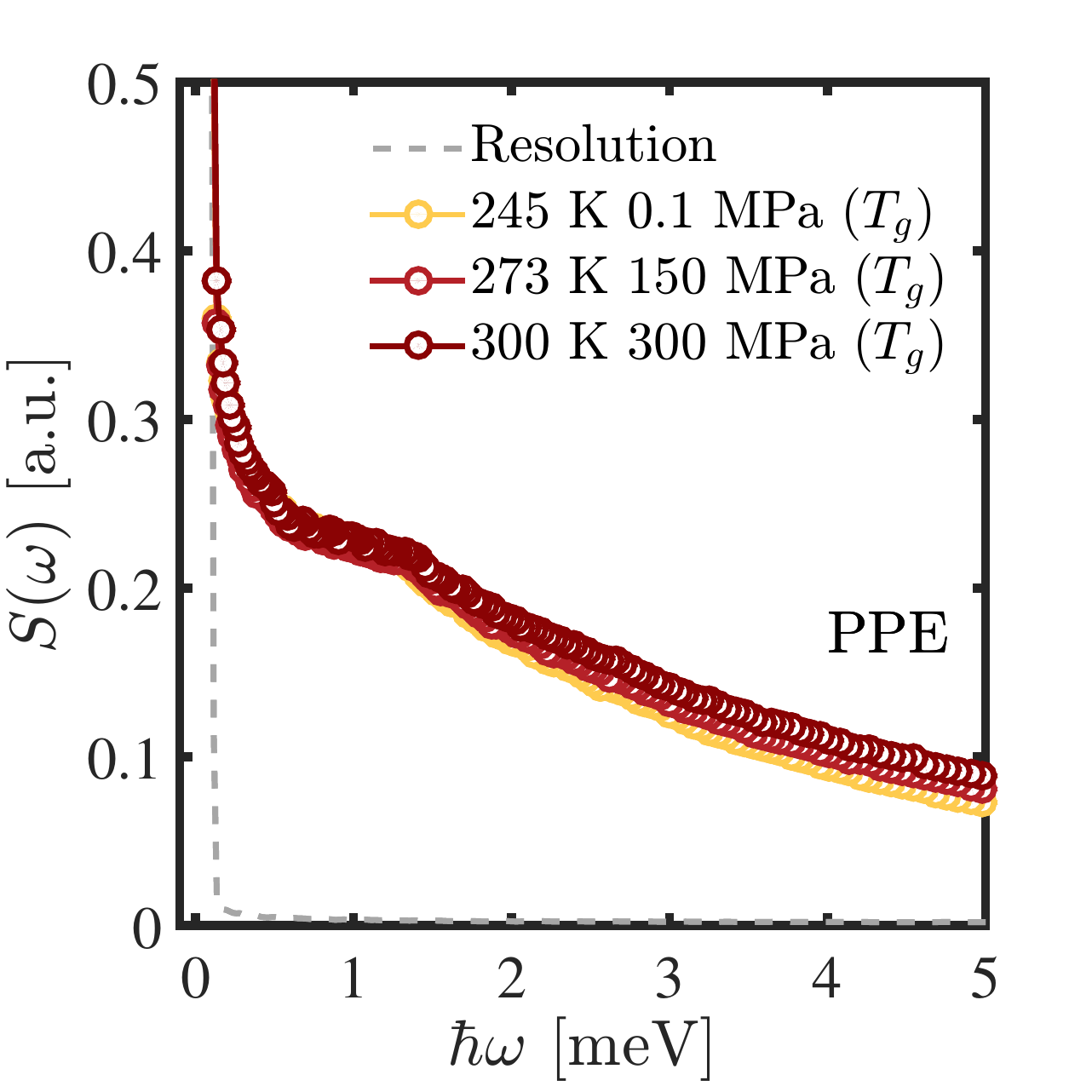} 
\hspace{0.1em}
\includegraphics[width=0.3\textwidth]{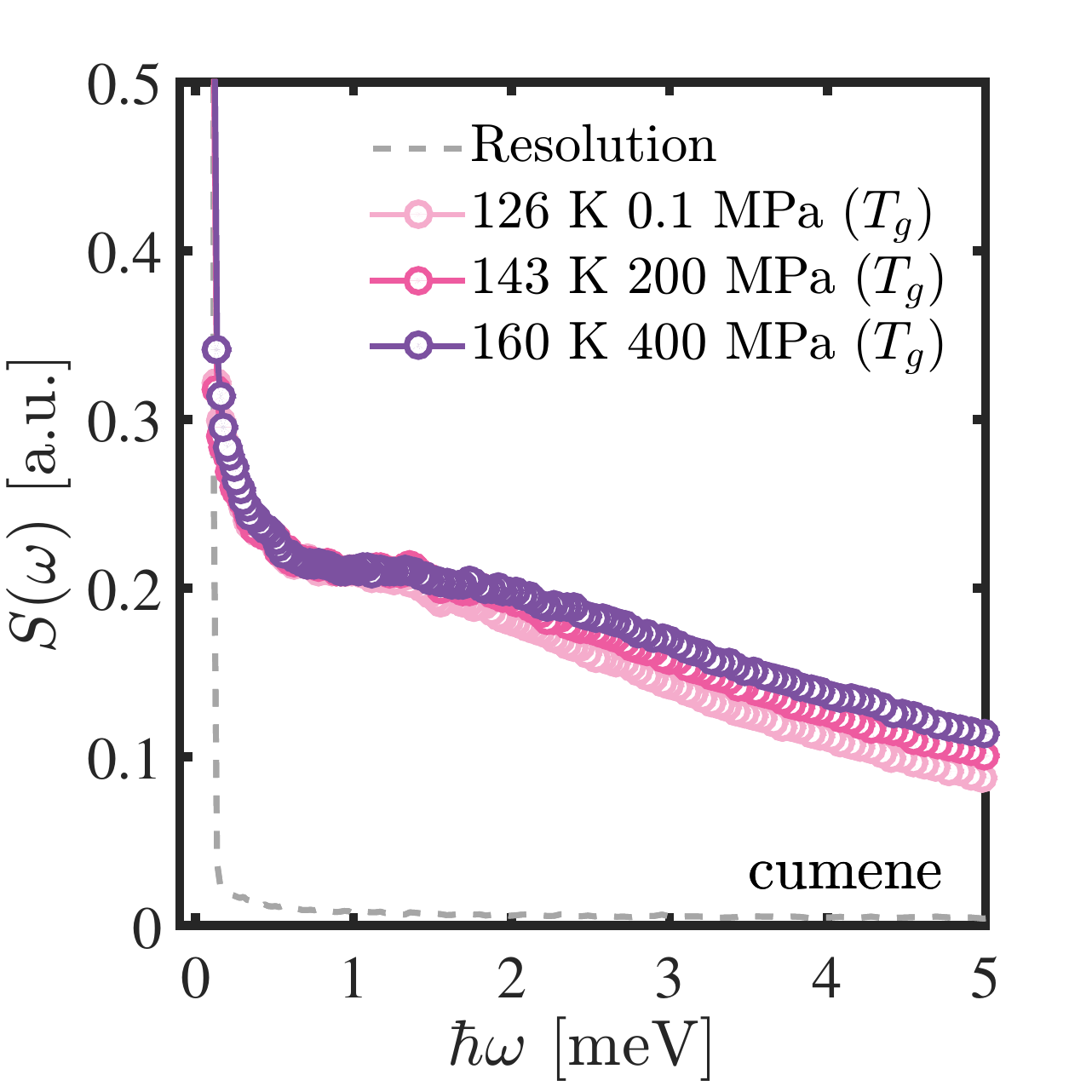} 
\hspace{0.1em}
\includegraphics[width=0.3\textwidth]{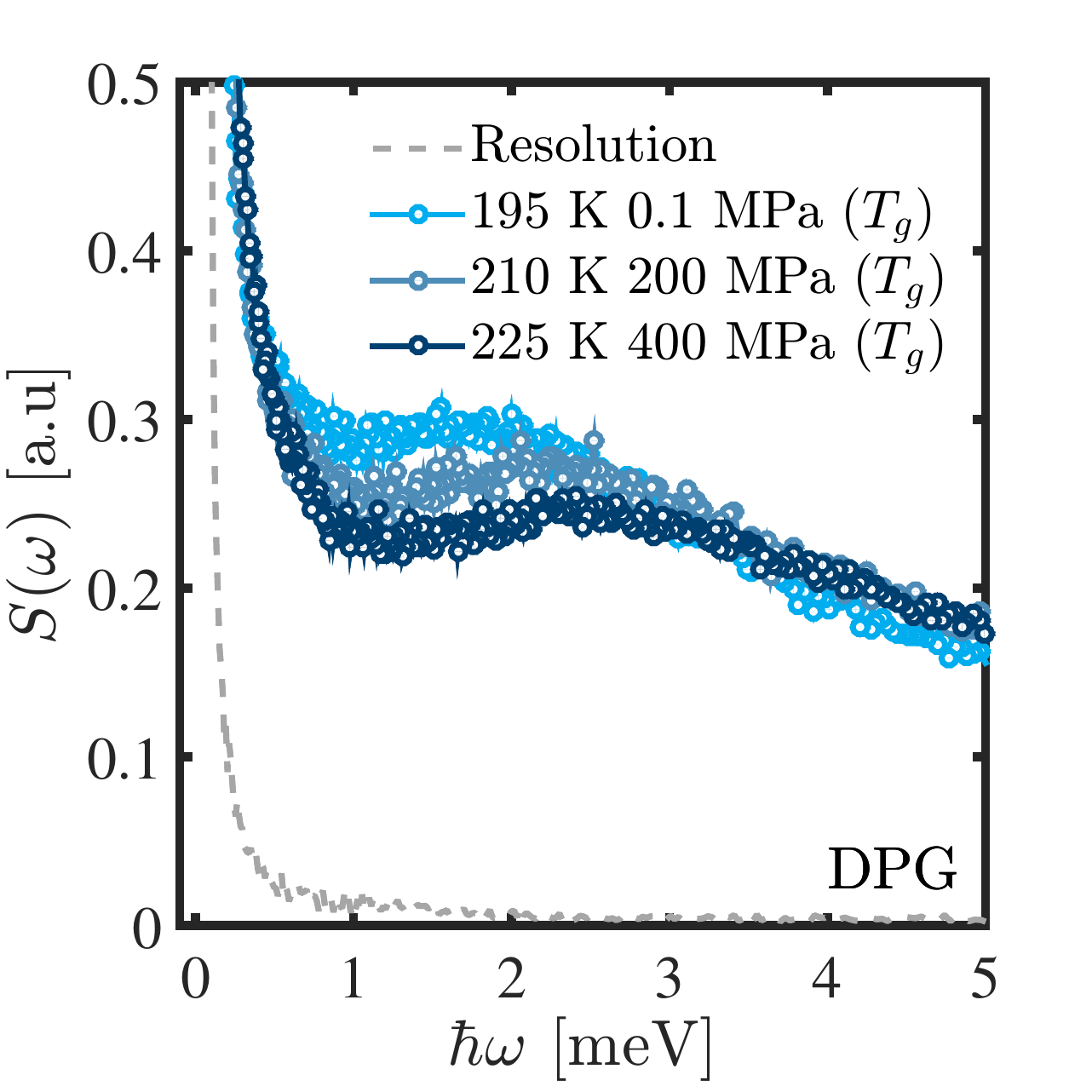}
\caption{Data from Fig.~2 on absolute energy scale. Picosecond dynamics at various state points for the three samples, PPE, cumene and DPG on absolute energy scale. Same state points and spectra as in Fig.~2, but here shown as measured on absolute energy scale.}
\end{center}
\end{figure}

\begin{figure}
\begin{center}
\includegraphics[width=0.4\textwidth]{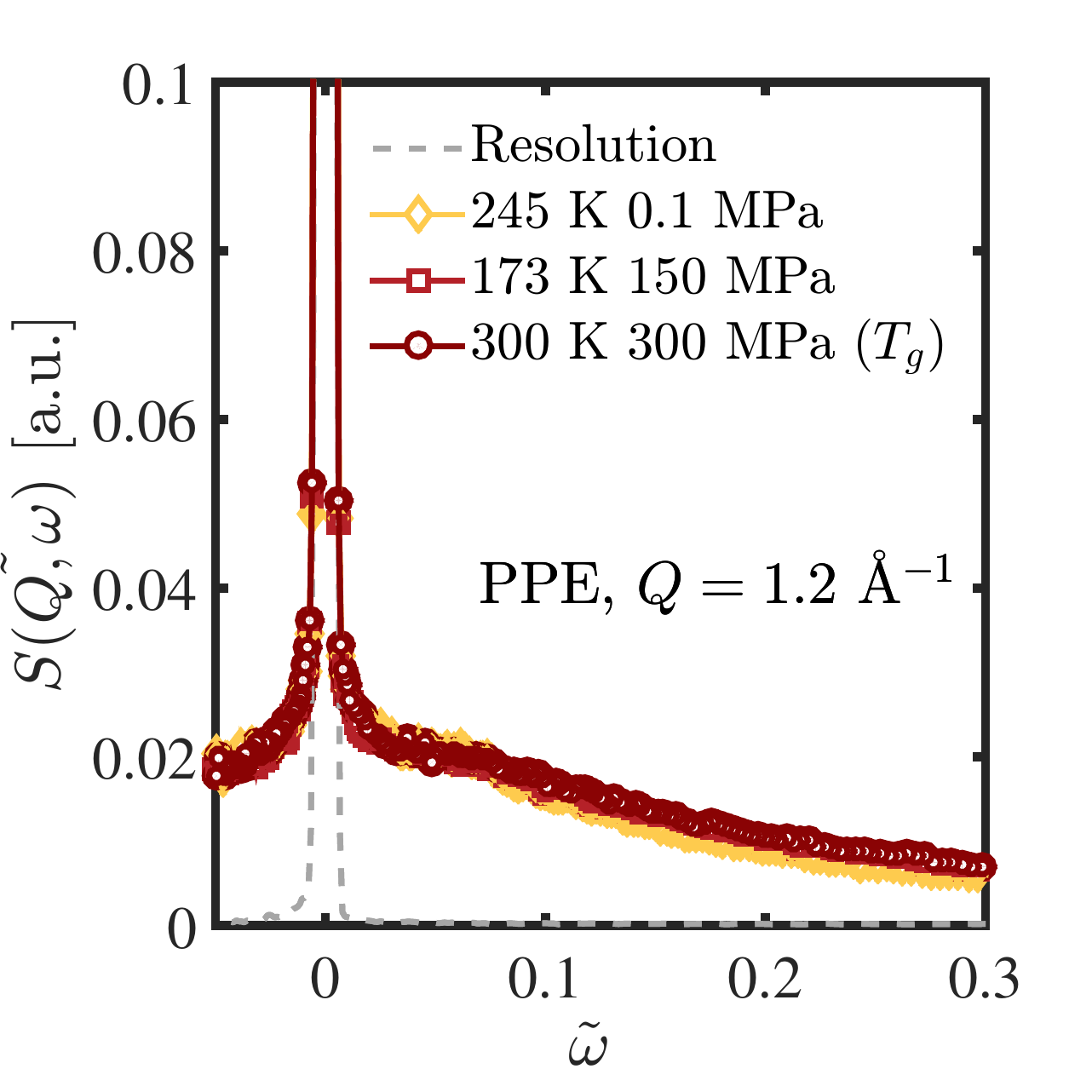}
\hspace{0.1em}
\includegraphics[width=0.4\textwidth]{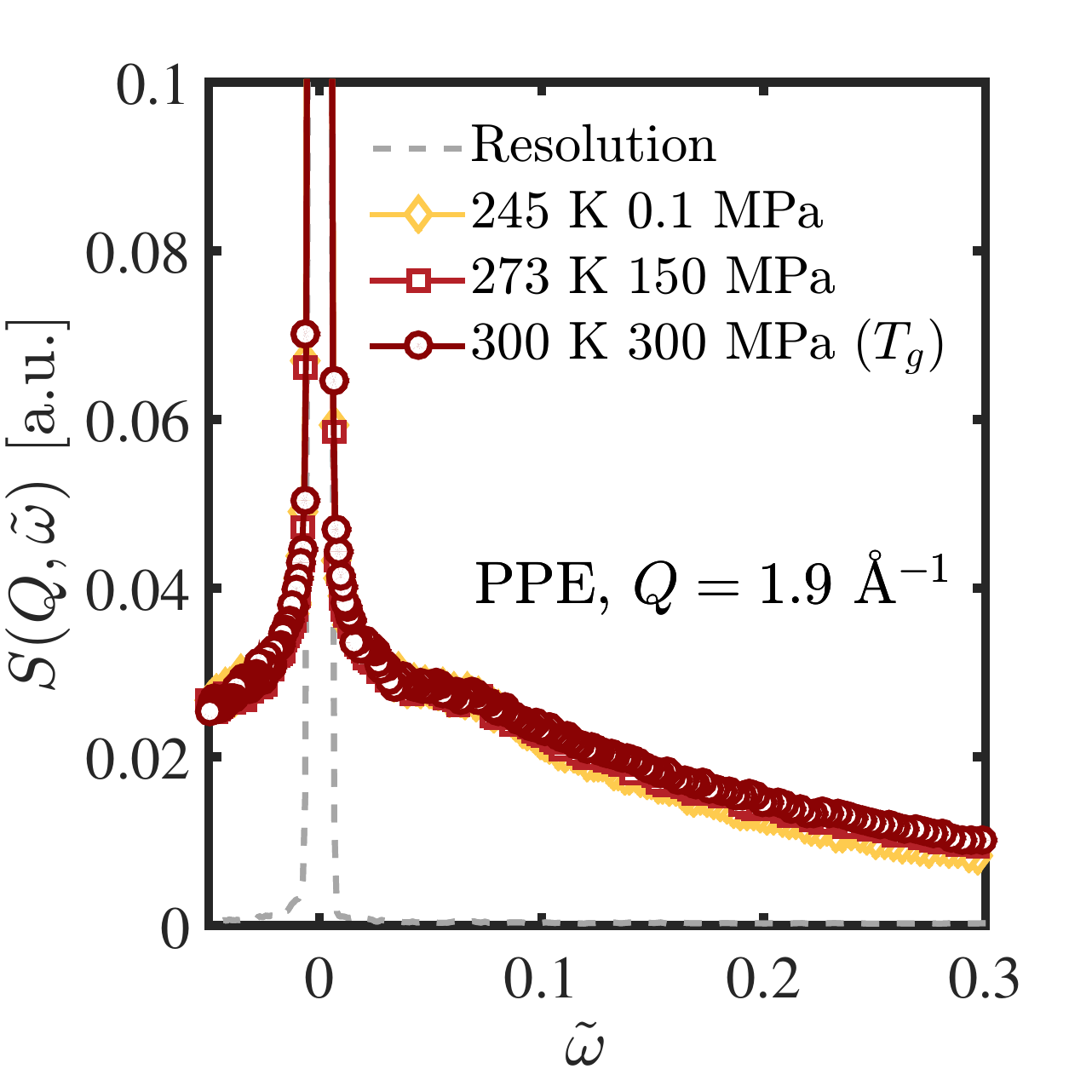}
\includegraphics[width=0.4\textwidth]{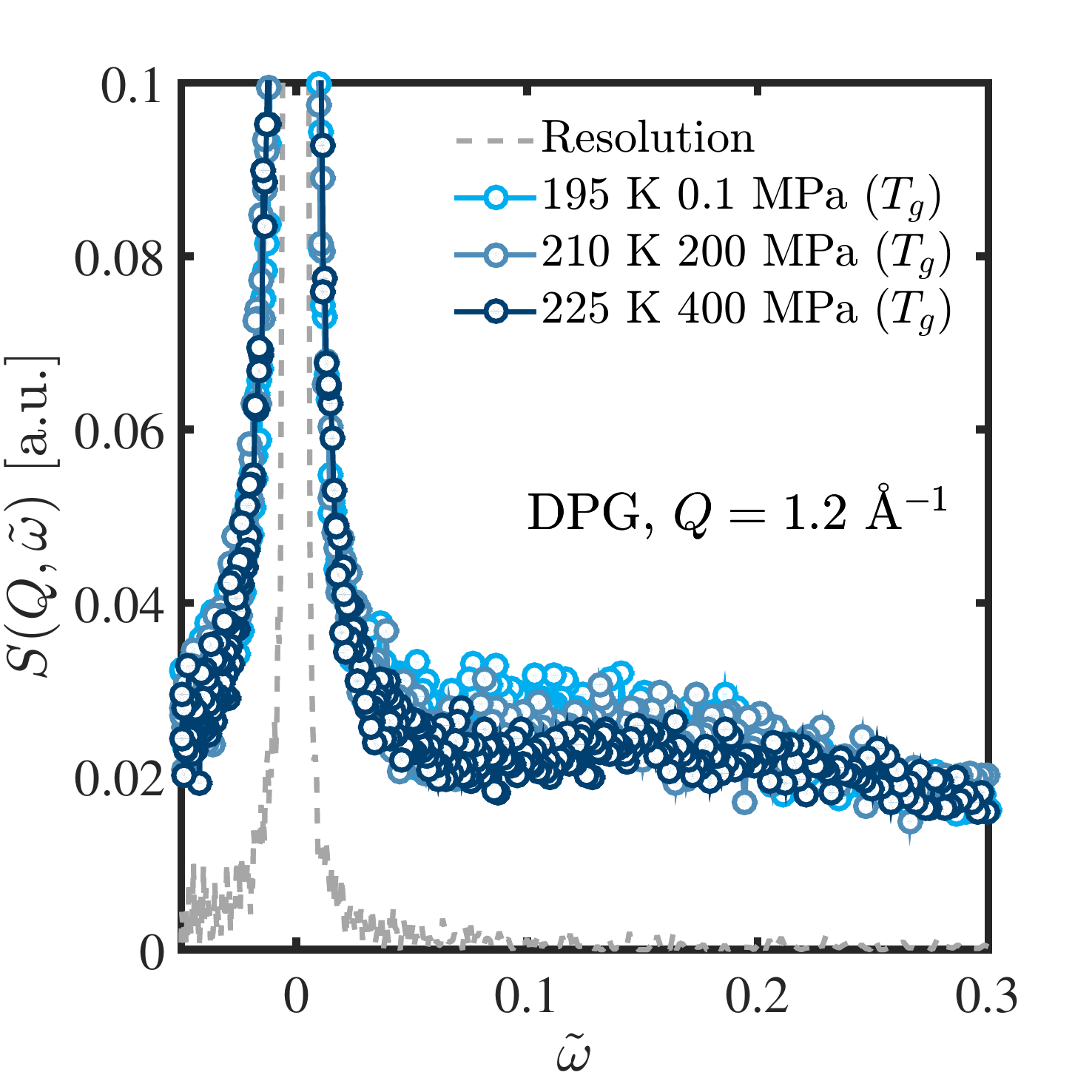}
\hspace{0.1em}
\includegraphics[width=0.4\textwidth]{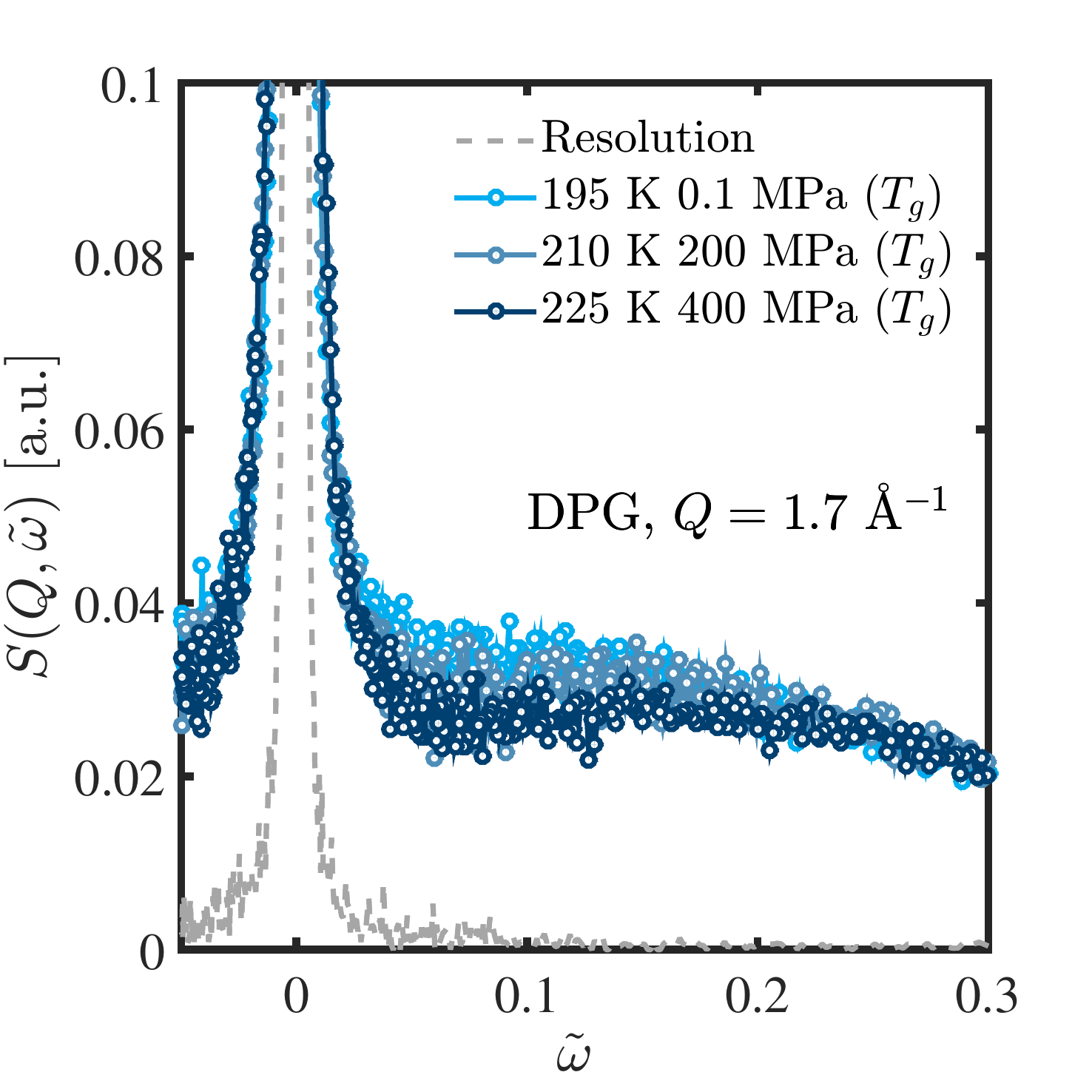}
\caption{An example of $Q$-dependence of spectra. Top: spectra of PPE. Bottom: spectra of DPG. Left: lowest value of $Q$. Right: higest value of $Q$. Same trend observed for all values of $Q$. Spectra plotted in reduced energy units $\tilde{\omega}=\omega\rho^{-1/3}T^{-1/2}$.}
\end{center}
\end{figure}

\begin{figure}[htbp!]
\centering
\textbf{a}
\includegraphics[width=0.47\textwidth]{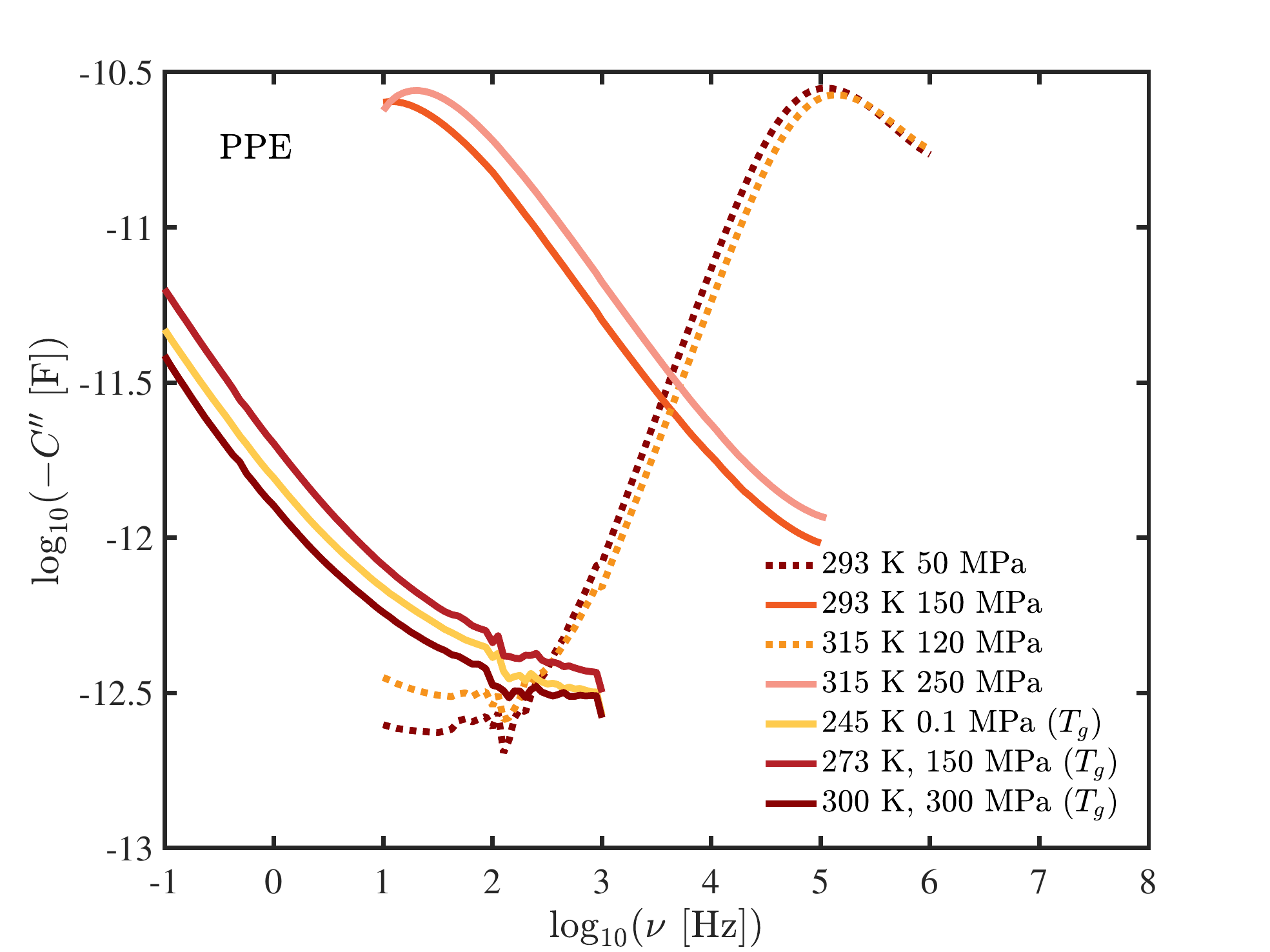} 
\textbf{b}
\includegraphics[width=0.47\textwidth]{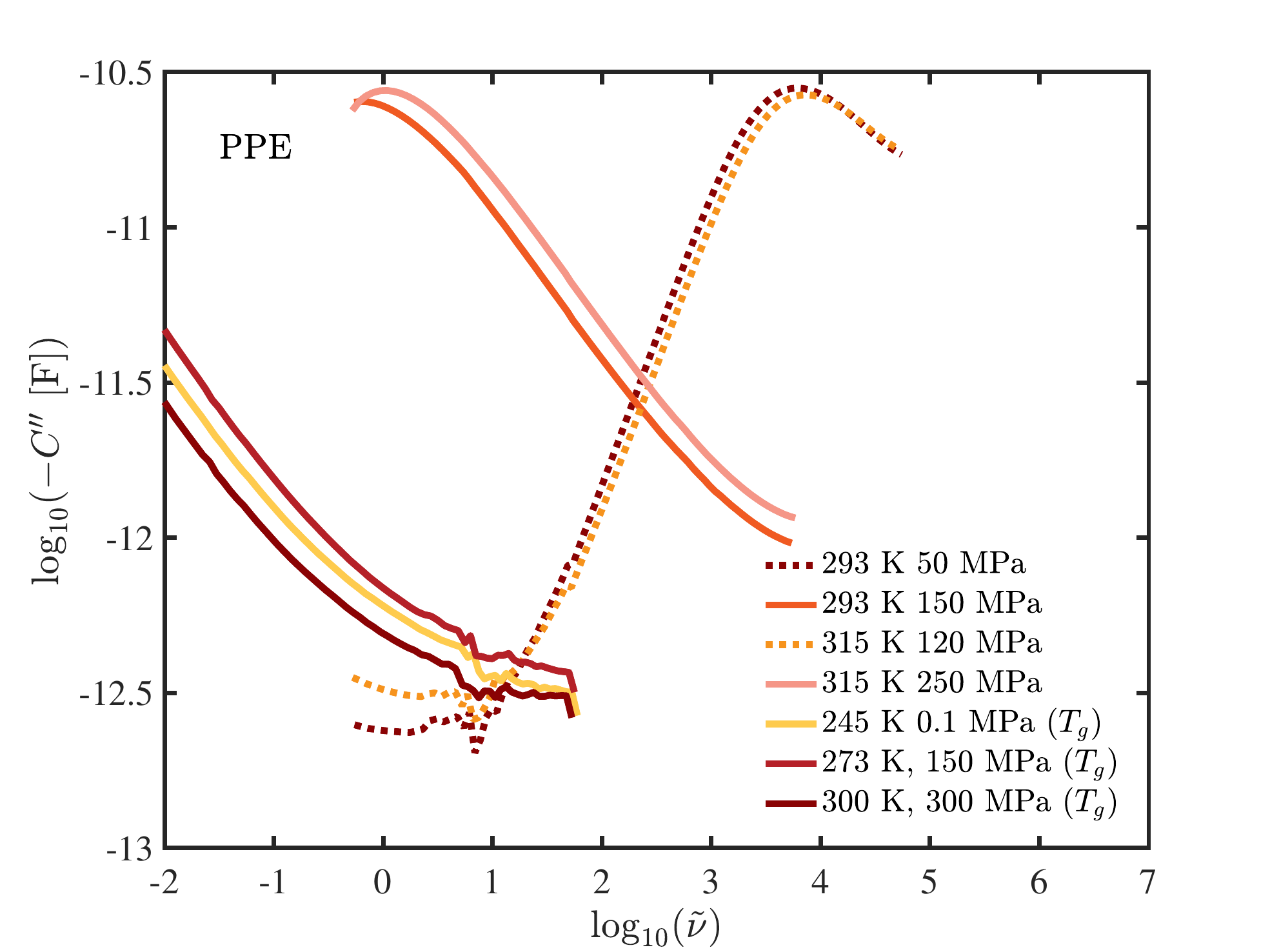}
\caption{Examples of isochrones found from dielectric spectroscopy. (a) Imaginary part of the capacitance on PPE along three isochrones, i.e. roughly the same alpha relaxation time $\tau_\alpha$ of temperature and pressure given by the maximum of the imaginary part of the capacitance. (b) Same as (a) but plotted in reduced units, $\tilde{\nu}=\nu\rho^{-1/3}T^{-1/2}$. Alpha relaxation time of the three isochrones: $\tau_\alpha\approx10^2$~s~$(T_g)$, $10^{-2}$~s, $10^{-6}$~s, respectively.}
\end{figure}

\begin{figure}[htbp!]
\centering
\textbf{a}
\includegraphics[width=0.47\textwidth]{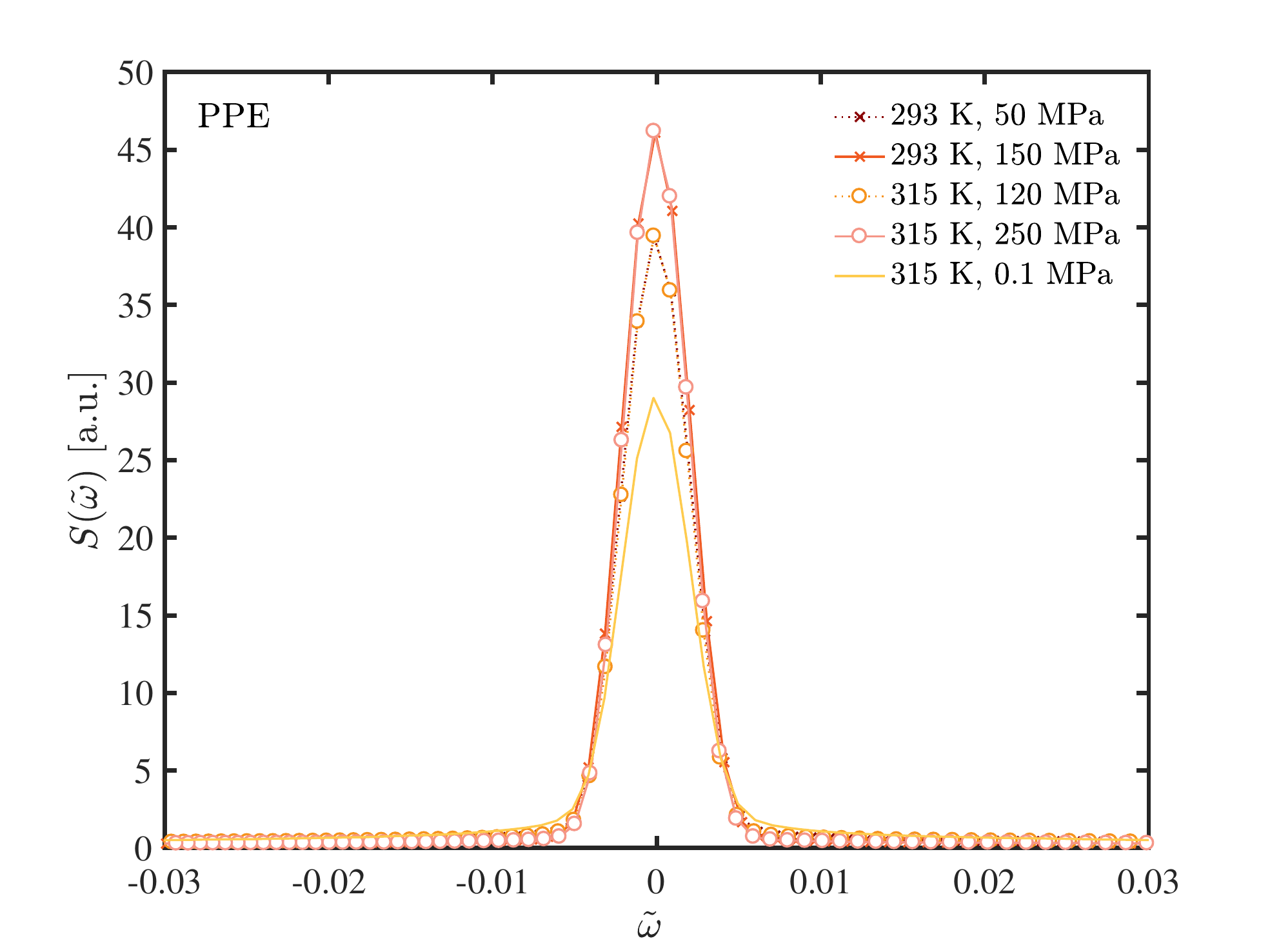}
\textbf{b}
\includegraphics[width=0.47\textwidth]{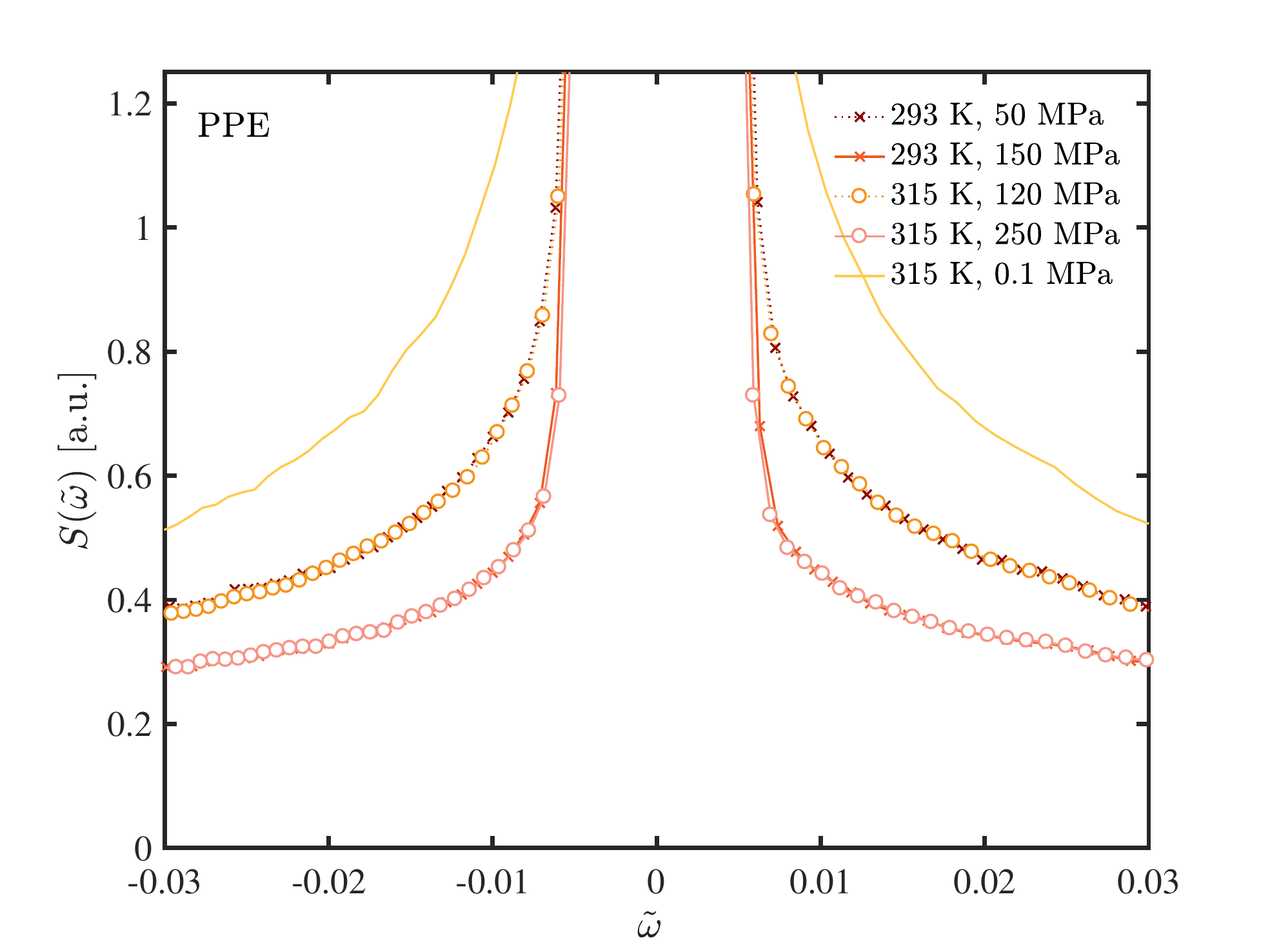}
\caption{Picosecond dynamics of higher temperature isochrones on PPE on IN5. Isochrones found from dielectric spectroscopy shown in Fig.~S3. (a) Spectra plotted in reduced energy units $\tilde{\omega}=\omega\rho^{-1/3}T^{-1/2}$ with $\tau_\alpha\approx10^{-2}$~s and $10^{-6}$~s. Full line for comparison, $\tau_\alpha\approx10^{-9}$~s. No scaling on $y$-axis of spectra. (a) Zoom of (b).}
\end{figure}

\end{document}